Title: **The use of quantum dots to amplify antigen detection**

Earl Bloch, PhD, Immunologist and Science consultant

Lou Massa, PhD,  Professor of Chemistry & Physics, Hunter College & the Graduate Center of the City University of New York, New York, NY 10016

Kilian Dill, PhD, Nanotechnologist and Science consultant

**Abstract:** Proposal to develop an Improved immunological assay employing primary IgG antibodies and secondary IgM antibodies labeled with quantum dots to amplify antigen detection.

**Corresponding authors:blochearl@gmail.com, lmassa@hunter.cuny.edu, redwoodranch@yahoo.com**

**Introduction:**

Rudolf Kraus in 1897 described the precipitation reaction from his studies of anti-cholera and anti-typhoid sera (1). Pauling in 1943 discussed the role of noncovalent forces (columbic, hydrogen bonding, hydrophobic, and Van Der Wall's) involved in antibody-antigen reactions that contribute to precipitation (2). Reverberi and Reverberi described factors affecting antibody antigen reactions in terms of the Law of Mass action and equilibrium constants (3). Marack in1934 proposed the Lattice theory which states "Specific precipitation of antibodies and molecular antigens would result from the same mechanism if both antibody molecules and antigen molecules were multivalent" (4). Based on Marack's theory, the antibody and its specific antigen must have at least two binding sites for effective cross linking to produce a visible lattice or precipitate (figure 1).

**Figure 1**

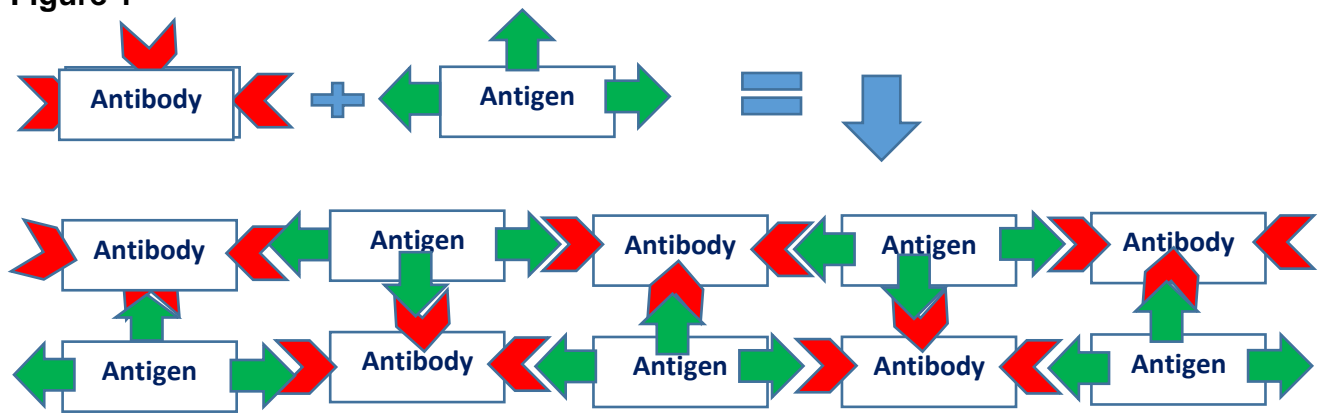

Alternatively, (univalent antibodies and univalent antigens), (multivalent antibodies and univalent antigens) and (univalent antibodies and multivalent antigens) will not result in extensive cross linking for lattice formation to occur.  In the late 1950s Soloman Berson and Rosalyn Yalow developed a very sensitive immunological assay that did not involve extensive cross linking of antibodies and antigens to form a visible precipitate for antigen detection. The investigators labelled only antibody molecules with radioactive iodine for the detection of insulin (antigen) concentrations in circulation (5). This assay was called a Radioimmunoassay or RIA. RIA is a member of a class of immunological assays that are called primary reactions. As stated primary reactions do not involve multivalent antibodies or multivalent antigens for lattice formation and antigen detection.

In primary reactions "visibility" occurs when the antibody is labelled with a radioactive element, an enzyme-fluorochrome or nanoparticles (quantum dots).

Precipitation is a secondary reaction and is not as sensitive as primary reactions because very high concentrations of the antibody and antigen are needed to form a visible lattice for antigen or antibody detection. Primary reactions do not involve extensive cross linking of antibodies and antigens and therefore are excellent for the detection of trace levels of antigens or antibodies in experimental and clinical samples.

The enzyme-linked immune sorbent assay (ELISA) was developed in 1971 to reduce the exposure to radioactive iodine. In the ELISA assay, the antibody is labelled with an enzyme-fluorochrome molecule (6). It should be clear that RIA, ELISA and quantum dots (QD) are not precipitation assays because optimal proportions of antibodies and antigen are not required for lattice formation. Quantum dots (QD) are fluorescence immunoassays that can amplify antigen detection at concentrations that are significantly lower than radioimmunoassay and ELISA. It should be noted that, QD have been used with success in numerous biological systems as well as in the of study viral antigens (7).

ELISAs can be classified as direct (primary antibody), indirect (primary and secondary antibodies), sandwich (two primary antibodies-first primary antibody serves to capture the antigen and the second primary antibody in labelled with an enzyme-fluorochrome to bind the captured antigen forming a "sandwich" for data collection (figure 2). Finally, the competitive assay involves two primary antibodies: unlabelled primary antibody competes with labelled primary antibody to bind a limited number of antigenic sites on a solid support. Additionally, the competitive ELISA can also involve labelled and unlabeled corticosterone (antigens) vying for limited number of sites on antibody molecules on a solid support shown below.

The direct, indirect and sandwich ELISAs have positive slopes for the determination of antigen or antibody concentrations. Conversely the competitive ELISA has a negative slope for the determination of antigen or antibody concentration. Positive slopes occur when the increasing concentrations of antibody bind to increasing concentrations of antigenic sites on a solid support (figure 3). Negative slopes occur when labelled and unlabeled primary antibodies compete for limited number of antigenic sites on a solid support (figure 4). In this approach, secondary IgM antibodies with quantum dots will produce a positive slope when it binds to a primary IgG antibodies in our experimental design (figure 3).

**Figure 2**

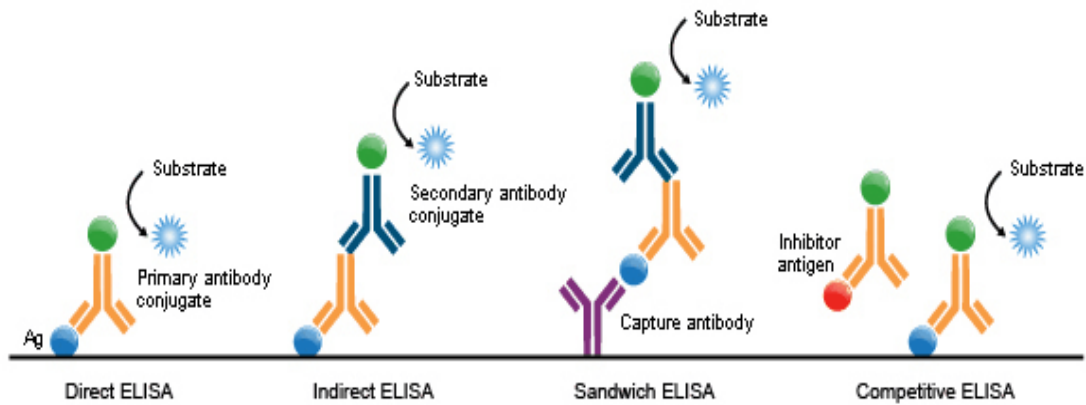

**Figure 3**

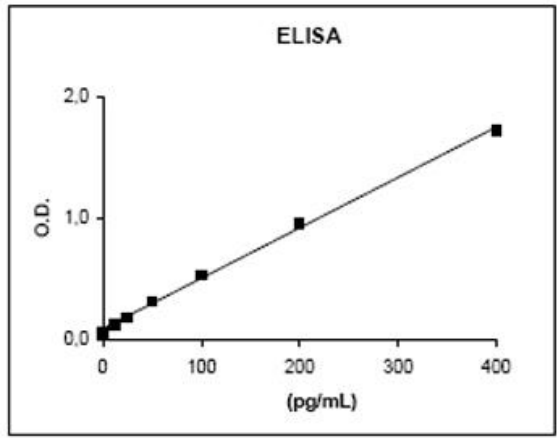

**Figure 4**

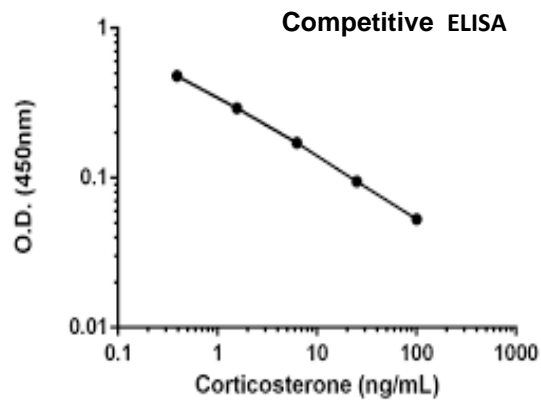

**Results:** A modified approach of Bloch, Massa and Dill, involves the use of two different isotypes or classes of antibodies IgG and IgM, see Fig. 1 below.  Monomeric IgG has two identical heavy chains, two identical light chains and **three constant regions on each heavy chain and one constant region on each light chain.**   IgM has five identical monomers. Each monomer has two identical heavy chains, two identical light chains and <u>**four constant regions**</u> **on each heavy chain and one constant region on each light chain.** With respect to structure, it should be clear that the IgM cyclic pentameric is not a polymer of five identical IgG molecules (see figure 5 below)

The variable regions on heavy and light chains defines the specificity of the antibody molecule for a given antigen.  It should be noted that both heavy chains and both light chains have identical sequences of amino acid residues in the variable regions in their respective chains.

For example, in the labelled figure of the IgG molecule (to the right) the amino acid residues on both heavy chains have the same exact sequence which we refer to "**QEZGPSTWYV" or the blue color.** The amino acid residues on both light chains have the same exact sequence that we have listed as **"ARNDBCQEMFPS" or the red color** (table 1). This illustrates that amino acid residues in the variable regions **of heavy and light chains** do not have identical sequences of amino acid residues to each other. Moreover it is important to note that each chain contributes 50% to the binding activity of the divalent antibody molecule. Finally, each antibody molecule to a different antigenic determinant will have its own unique amino acids residues in the variable regions on both heavy chains. This argument also applies to the light chains as well. That is, the sequence of amino acid residues in the variable regions of both light chains are identical to each other but completely different from those in the variable regions for both heavy chains.

Table 1

Abbreviations for amino acids

| Amino acid | Three-letter abbreviation | One-letter symbol |
|---|---|---|
| Alanine | Ala | A |
| Arginine | Arg | R |
| Asparagine | Asn | N |
| Aspartic acid | Asp | D |
| Asparagine or aspartic acid | Asx | B |
| Cysteine | Cys | C |
| Glutamine | Gln | Q |
| Glutamic acid | Glu | E |
| Glutamine or glutamic acid | Glx | Z |
| Glycine | Gly | G |
| Histidine | His | H |
| Isoleucine | Ile | I |
| Leucine | Leu | L |
| Lysine | Lys | K |
| Methionine | Met | M |
| Phenylalanine | Phe | F |
| Proline | Pro | P |
| Serine | Ser | S |
| Threonine | Thr | T |
| Tryptophan | Trp | W |
| Tyrosine | Tyr | Y |
| Valine | Val | V |

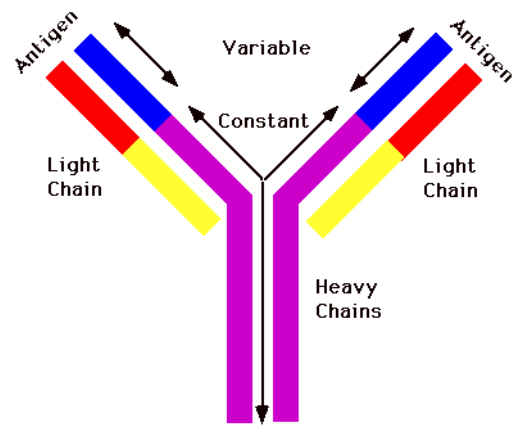

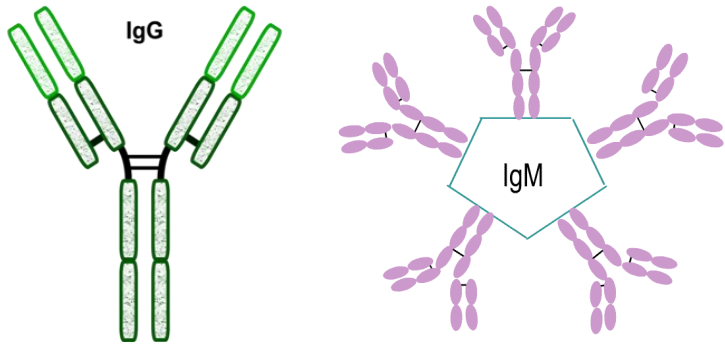

**Figure 5**

A representation of IgG and IgM molecules.

IgG is the primary antibody in our novel approach and IgM is the secondary antibody. By definition a primary antibody (IgG) has specificity for a given antigenic determinant. The secondary IgM antibody has specificity for a given primary (IgG) antibody (figure 6).

**Figure 6**

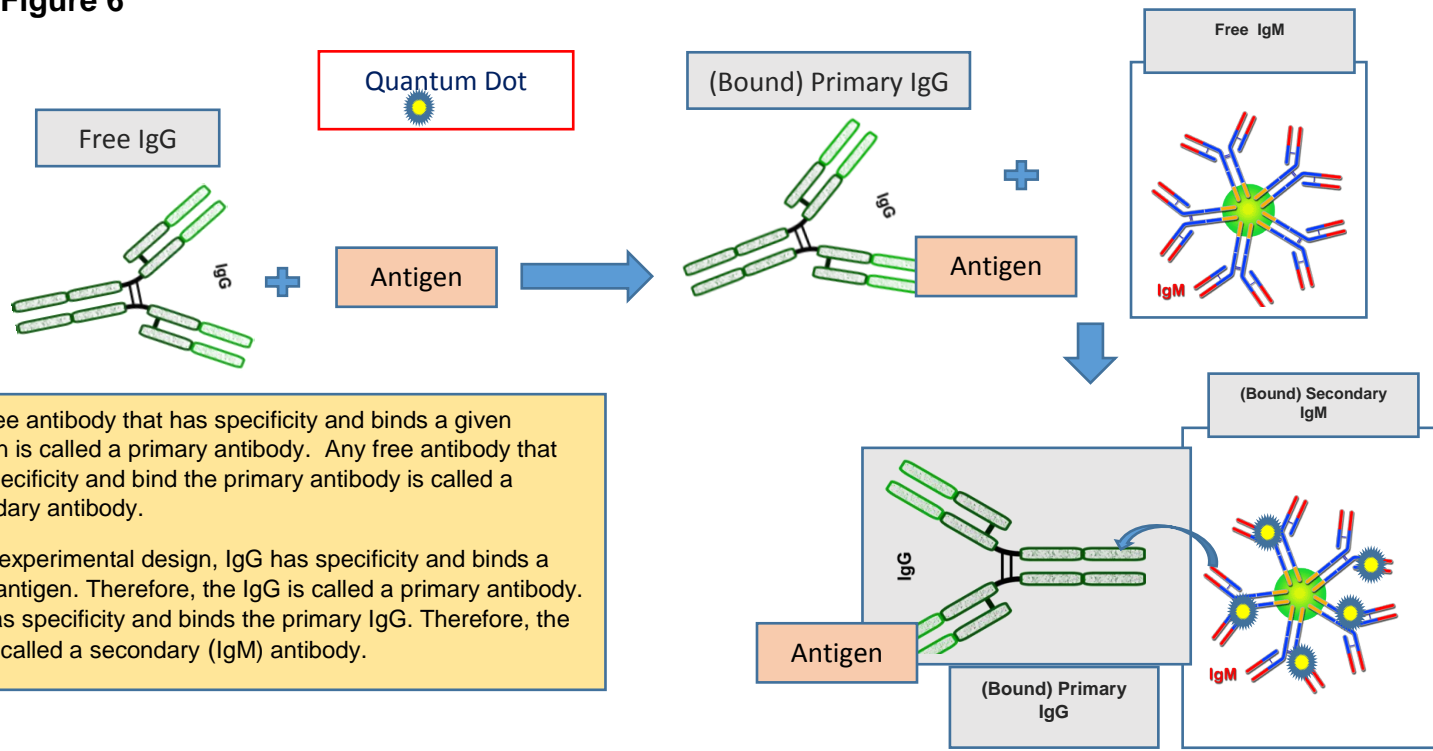

Any free antibody that has specificity and binds a given antigen is called a primary antibody. Any free antibody that has specificity and bind the primary antibody is called a secondary antibody.

In our experimental design, IgG has specificity and binds a given antigen. Therefore, the IgG is called a primary antibody. IgM has specificity and binds the primary IgG. Therefore, the IgM is called a secondary (IgM) antibody.

Researchers (6) have labelled IgM antibodies with quantum dot in their immunological assays. However, in the article cited below the IgM was used as a primary antibody not a secondary antibody (8). In our approach IgM is labelled with quantum dots and used as a secondary antibody to bind to primary IgG antibodies to amplify and enhance signal detection.

**How to produce a specific human secondary IgM**

Secondary IgM antibodies that are specific for IgG can be produced in at least two ways.

Method 1. Inject human IgG or its heavy chains into experimental animals such as rabbits and after 10 days to two weeks collect the serum samples and recover the IgM fraction and determine its titer. Label the recovered IgMs fraction with quantum dots using a published protocol (9).

Method 2. Dissociate IgM from rheumatoid factor (10), determine its titer and label the dissociated IgM with quantum dots (9,).

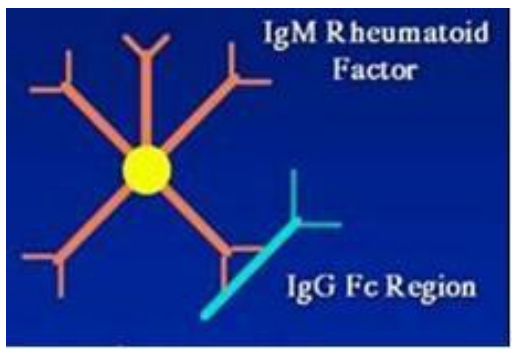

**Figure 7.** Secondary human IgM Rheumatoid antibody binds primary IgG molecule.

Investigators at Thermo Fisher Scientific have developed unique system for linking markers such as enzymes, quantum dots and number of ligands to antibody molecules for antigen detection. It is called siteclic antibody labelling system. Basically, it involves the chemical linking of a desired marker to side chain carbohydrates on antibodies molecules. We have modified their method to label secondary IgM antibodies for use in our proposal because of the numerous number of carbohydrates that are found in cyclic pentameric IgM.

**Monomeric IgM has significantly more carbohydrates sites then monomeric IgG, ten vs two respectively (figures 8, 9). IgM exist predominately as a cyclic pentamer** with five monomer units with a total of **50 carbohydrates sites. In our novel approach we can use the siteclick method to increase the covalent linkage of quantum dots to the carbohydrates sites in the IgM antibody molecule. It is anticipated that this procedure will increase the sensitivity of our secondary IgM antibodies to bind any primary IgG antibodies and specific antigens in trace levels in clinical and experiment samples as detailed above.**

Figure 8              Figure 9

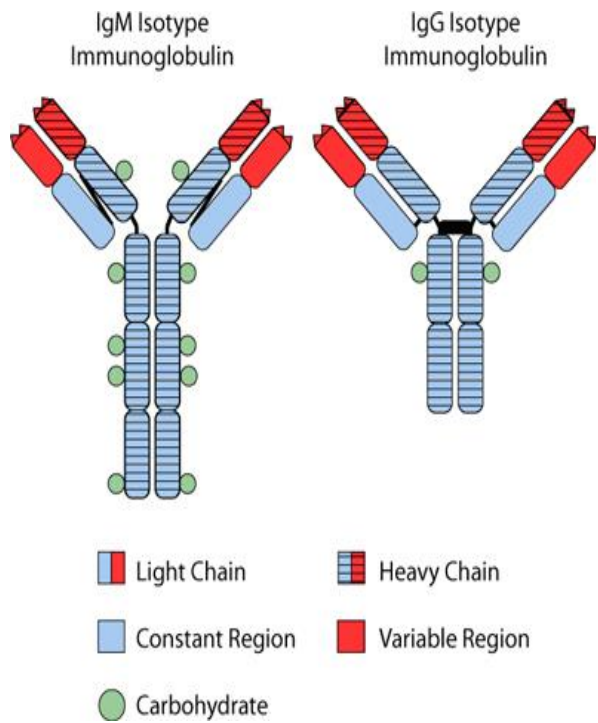
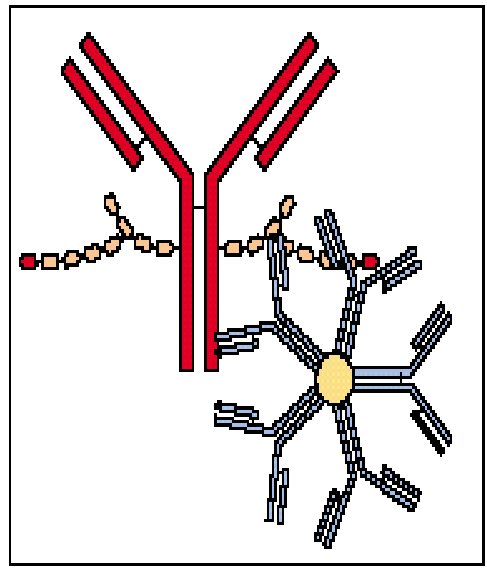

**Discussion:**

In the early 1980s. Alexei Ekimov (11), working at the Vavilov State Optical Institute, in ST. Petersburg Russia investigated semiconductor nanocrystals, or colloidal quantum dots. Dr. Ekimov discovered the presence of nanosize crystals of these semiconductors in the glasses and that these nanocrystals could absorb light at much lower wavelengths than expected. Nanocrystals or quantum dots consist of 10-50 atoms and the nanocrystals can produce distinctive colors via fluorescence was determined by the size of the particles (12). This was an excellent discovery because microarrays or biochips with attached peptides for proteins analysis, drug screening, immunological assays, were usually visualized and analyzed using organic fluorescent dyes (8). Unfortunately the dyes had poor photo stability, and decreasing quantum yields upon conjugation to biomolecules.

Based on numerous and documented successes with quantum dots, these procedures are replacing Enzyme linked immune assays (ELISA). It should be clear that our novel quantum dot assay employs secondary IgM antibodies to bind primary IgG molecules to amplify antigen detection.

It should be clear that current Quantum dot (QD) assays do not use secondary IgM antibodies as the indicator system. They use primary labelled IgG (or IgM) as the indicator system. Problems with this approach. For example, if 100 unrelated primary IgGs are involved in a hypothetical study, each of those 100 primary IgGs would have to be labelled with QD for specific detection of a given antigen. In our approach, we do not label any of the 100 primary IgGs in the above hypothetical study. We simplify the procedure by labelling secondary IgM antibodies with quantum dots to identify all of the 100 primary IgG antibody molecules. This works because the secondary IgM will bind to sites that are common to all IgG antibody molecules

**Procedure for the hypothetical study.**

Add each of the100 unique and unrelated antigens to 100 separate labelled wells in microtiter plates for incubation with its specific primary IgG. After the incubation period rinse each well to remove unreacted primary IgG. Now add the labelled secondary IgM with quantum dots to each of the 100 wells for the second incubation period. Finally, rinse the wells to remove unreacted secondary IgM labelled antibodies and analyze the data. It should be clear that the secondary IgM will recognize the 100 primary IgGs in the separate wells in the microtiter plates that are bound to each specific antigen. As stated above this works because the secondary IgM is bind to sites that are common to all IgG molecules. Therefore, secondary IgM "does not care" what specific antigens primary IgGs recognizes -all that concerns the secondary IgM is the presence of common sites or epitopes on all primary IgG antibodies for secondary IgM to bind.

However, if the primary IgG antibodies did not recognize the antigen in the first incubation step, the primary IgG will not be bound and would be washed away when the wells are rinsed. Therefore, when the secondary IgM is added for the second incubation step- it cannot bind because the primary IgG antibodies were not present. Finally, the wells are washed to remove unreacted secondary IgM before data collection.

Malito, Carfi and Bottomley have written an excellent review on the use of protein crystallography in vaccine research development.  Additionally, they described how structural information and protein engineering can lead to the development of antigens that are safe, protective and easy to develop(13).  The kernel energy method (KEM) or quantum crystallography (14-16) in which Lou Massa is a co-developer can apply quantum mechanics to study peptides with the best negative interaction energies as potential candidates for vaccine development. The kernel energy method (KEM) carried to second order has been used to calculate the quantum mechanical *ab initio* molecular energy of peptides, protein (insulin and collagen), DNA, and RNA and the interaction of drugs with their biochemical molecular targets (17).  The novel quantum dot approach in this proposal will allow us to monitor antibody response to the best peptides from bacterial and viral pathogens that were chosen from KEM calculations.  Our immunological assay involving secondary IgM antibodies amplifies antigen and antibody detection and reduces the cost and time for investigators in the laboratory to obtain such data. Finally, we are seeking collaboration with technology companies to market our unique approach.